# Local distortion of MnO$_6$ octahedron in La$_{1-x}$Sr$_x$MnO$_{3+\delta}$ (x = 0.1 – 0.9): An EXAFS study

R. Bindu, S. K. Pandey, Ashwani Kumar[a], S. Khalid[b], and A. V. Pimpale

UGC-DAE Consortium for Scientific Research, University Campus, Khandwa Road, Indore 452 017, India

[a]School of Physics, Devi Ahilya University, Khandwa Road, Indore 452 017, India

[b]National Synchrotron Light Source, Brookhaven National Laboratory, Upton, New York – 11973

E-mail: sk_iuc@rediffmail.com

## Abstract

Room temperature Mn K-edge extended X-ray absorption fine structure (EXAFS) studies were carried out on La$_{1-x}$Sr$_x$MnO$_{3+\delta}$ (x = 0.1 – 0.9) compounds. It is found from the detailed EXAFS analysis that the local structure around Mn sites is different from the global structure inferred from X-ray diffraction, especially for x ≤ 0.4, indicating presence of local distortions in MnO$_6$ octahedra. For the rhombohedral compounds, x = 0.1 to 0.3 the distortion is maximum for x = 0.1 and two bond lengths are seen- short one in basal plane and long one in apical plane. For compounds with x = 0.4 to 0.8 two short bonds in basal plane and four long bonds- two in the basal plane and remaining two in the apical plane are seen. For the compounds up to x = 0.3 compositions long bond length decreases and short bond length increases with increase in x whereas for the compounds 0.4 ≤ x ≤ 0.8 both types of bond lengths decrease. Such behaviour of bond lengths is an indication of the changed nature of distortion from Jahn-Teller type to breathing type at x = 0.4 composition.

# 1. Introduction

Transition metal oxides show interesting behaviour like metal insulator transition, colossal magneto resistance (CMR), charge ordering (CO), orbital ordering (OO), magnetic ordering etc. Currently, in this field, much research activity [1-5] is focused on manganites belonging to the perovskite family as they show CMR effect. The compounds under study $La_{1-x}Sr_xMnO_{3+\delta}$ also show above interesting properties for specific ranges of x. Over composition range $0.175 \leq x \leq 0.5$ transition from a paramagnetic insulator to a ferromagnetic metal is seen as a function of temperature. The coexistence of ferromagnetism and metallicity was explained earlier in terms of the double exchange (DE) model, however more recently Millis *et al.* [6] have pointed out the inadequacy of DE mechanism in explaining the observed properties and proposed a mechanism based on polaron formation. The doping of Sr at La sites causes two-fold effects: (1) to vary the number of electrons (band filling) and hence change the electronic configuration, and (2) the size effect, which changes the inter-atomic distances and bond angles. Depending upon the amount of chemical pressure introduced into the lattice, changes occur in the local environment and later on extend throughout the lattice. These changes are reflected in the Mn-O bond distances and Mn-O-Mn bond angles, which play crucial role in governing the physical properties. Thus Mn K-edge X-ray Absorption Spectroscopy (XAS) studies would help in understanding the above effects.

EXAFS region starts from about 50 eV above the edge and extends typically up to about several hundred electron volts. In this region, the photoelectrons have high kinetic energy ($E-E_0$ is large), and single scattering by the nearest neighbouring atoms normally dominates. The fine structure observed in this region is mainly because of interference of the ejected photoelectron represented by the outgoing wave and the back-scattered wave from the neighbouring atoms [7-9].

Many workers [10-15] have studied local structure around Mn ions in manganites with varying temperature and magnetic field with the aim of understanding magneto-transport properties. Extensive EXAFS study has been carried out for $La_{1-x}Ca_xMnO_3$. In the case of $La_{1-x}Sr_xMnO_3$, reports on local structure studies are available mainly for lower doped samples up to $x < 0.5$. Louca *et al.* [16] studied neutron diffraction pair distribution function (PDF) of $La_{1-x}Sr_xMnO_3$ ($0 \leq x \leq 0.4$) at T = 10 K and 300 K. In the rhombohedral metallic phase they have observed distortion in $MnO_6$ octahedra for compounds up to x = 0.3 compositions and attributed it to formation of Jahn-Teller (JT) polaron. Billinge *et al.* (for $La_{1-x}Ca_xMnO_3$; x = 0.12, 0.21 and 0.25 compounds) [17] also observed distortion in the $MnO_6$ octahedra and attributed it to breathing type. Hibble *et al.* [18] carried out PDF by pulsed neutron diffraction and have reported no such distortion in $La_{0.7}Sr_{0.3}MnO_3$ below $T_c \approx 370$ K, i.e. in metallic phase. Mellergard *et al.* [19] have reported absence of local distortion below $T_c$ for x = 0.2 sample using neutron diffraction and inverse Monte Carlo analysis. They observed distortion above $T_c$ attributing its link with $Mn^{4+}$ ion, which is different from JT type associated with $Mn^{3+}$ ion. They remark that the second Mn-O peak observed by Louca *et al.* [16] may be an analysis artifact. Shibata *et al.* [20] have carried out EXAFS study on $La_{1-x}Sr_xMnO_3$ ($0 \leq x \leq 0.475$) for two temperatures 10 K and 300 K, in the photoelectron wave vector range 3-15 Å$^{-1}$. They have also observed distribution in Mn-O bonds in rhombohedral structure. The difference in long and short bond lengths is small in comparison to that observed by Louca *et al.* [16]. Recently, Mannella *et al.* [21] have also carried out XAS and XPS studies on these compounds for x = 0.3 and 0.4 compositions and shown the presence of JT distortion. They have shown correlation between the disappearance of the splitting in the O K-edge pre-edge region and the presence of JT distortion.

It seems that the presence or absence of distortion in the $MnO_6$ octahedra in rhombohedral metallic ferromagnetic phase where double exchange is expected to be the dominant mechanism is still an open question. Moreover, for $x \geq 0.5$ compositions although the phase diagram reveals interesting behaviour [22] little EXAFS work is seen in the literature. The present work on Mn K-edge EXAFS covering the composition range 0.1 to 0.9 would thus yield information about the nature of distortion produced in $MnO_6$ octahedra in this system. We present here, the room temperature (T = 293 K) EXAFS studies for the first coordination shell for all the compounds. The compounds show structural transitions from perovskite to layered type and rhombohedral to orthorhombic with Sr doping [22,23]. We have seen distortion in the $MnO_6$ octahedra over the entire composition range covering different structures. For compounds up to x = 0.3 compositions with rhombohedral structure there are four short Mn-O bonds in the basal plane and two long in the apical plane. For $0.4 \leq x \leq 0.8$ compositions two short Mn-O bonds in the basal plane and four long Mn-O bonds, two of which in the basal plane and remaining two in the apical plane are seen. Here the x = 0.4 compound is rhombohedral and rest are orthorhombic. In the end layered compound there are three short and three long Mn-O bonds. The nature of distortion is of JT type for x < 0.4 and of breathing type for x ≥ 0.4.

**2. Experiment and data analysis**

The details of sample preparation and characterization are given in our earlier publication [23]. In short, powder samples of $La_{1-x}Sr_xMnO_{3+\delta}$ (x = 0.1- 0.9 in steps of 0.1) were prepared by solid-state reaction of $La_2O_3$, $SrCoO_3$ and $MnO_3$ with repeated grinding and calcinations at 1000 °C. Final sintering for all the samples was done at 1400 °C for two days to have better crystalline quality. All the samples were characterized by X-ray powder diffraction at room temperature and found to be single phase. The Rietveld analysis of diffraction patterns revealed that the crystal structure is rhombohedral for 0.1 ≤ x ≤ 0.4, orthorhombic for 0.5 ≤ x ≤ 0.8, and layered for x = 0.9 compositions. Iodometric redox titration was also carried out to estimate the oxygen nonstoichiometric δ, which is given in Table 1. From the Table it is evident that while the compounds x ≤ 0.4 are oxygen nonstoichiometric, the compounds 0.5 ≤ x ≤ 0.9 are nearly stoichiometric within the experimental accuracy.

Room temperature Mn K-edge XAS experiments were done at beamline X-18 B at the National Synchrotron Light Source, Brookhaven National Laboratory. The storage ring was operated at 2.8 GeV, 300 mA. The beamline used a Si (111) channel cut monochromator. The horizontal acceptance angle of the beam at the monochromator was 1 mrad. The vertical slit size used in this experiment was 1 mm, corresponding to an energy resolution of 0.8 eV at the Mn K-edge. The average photon flux for this bandwidth was $10^{10}$ photons/sec. The monochromator was detuned by 35 % to reduce the higher harmonics. The incident ($I_0$) and the transmitted beam ($I_t$) were measured by sealed ion chambers, with the combination of gases for appropriate absorption. Standard Mn foil was placed between the detectors $I_t$ and $I_{ref}$ for energy reference and to check the stability of the beamline and optical system. The samples sieved through a 400 mesh were spread uniformly on a cellophane tape and a four-fold of this tape was used to minimize the pinhole and brick effects.

EXAFS fitting was carried out by using UWXAFS 3.0 software [24]. The threshold energy, $E_0$, for all the spectra was taken as the first inflection point in the absorption edge region. After the background subtraction, the absorption coefficient $\mu(E)$ was converted to $\mu(k)$, where $k = (2m(E-E_0)/\hbar^2)^{1/2}$ is the magnitude of wave vector of the ejected photoelectron. The XAFS oscillation $\chi(k)$ is defined as, $(\mu-\mu_0)/\mu_0$, where $\mu_0$ is the embedded atom absorption coefficient. The Fourier transform (FT) to the r-space was taken in the k range 3-13 $Å^{-1}$ by Fourier transforming $k^2\chi(k)$ with Hanning window. First shell fitting was done in the Fourier filtered k space in the range 0.76 to 1.93 Å for all the compounds $x \leq 0.8$. For $x = 0.9$ composition, the EXAFS fitting was carried out in the range 0.89 to 1.93 Å. The upper limit of the filtering window was chosen by checking the FT of the theoretical $\chi(k)$ calculated using FEFF6.01 [25], in which the peaks related with the $MnO_6$ octahedron were observed only below the upper limit indicated above. The overall many body reduction factor $S_0^2$ was fixed to 0.82 for all the samples. The back scattering amplitude and phase shifts were calculated using FEFF6.01 for $LaMnO_3$ and the same were used for all the compounds. During the fitting N was kept fixed as per the choice of model structure and only $\sigma^2$ was varied to reduce the number of correlated parameters. Further, several different fits were performed in each case to verify the robustness of the parameters. Both- simultaneous variation of different fitting parameters as well as their independent variation was tried. A fit is considered to be good if the goodness of fit parameter given by R-factor is less than 0.02 [24]. Since for every composition fitting was tried using different structure models, the model that resulted in the least R-factor was considered to represent the local structure for that particular composition.

We have used different model structures in order to get the information about the local structure in these compounds. The average structure inferred from X-ray diffraction for the compounds up to $x = 0.4$ compositions is rhombohedral [23]. Therefore, one expects that all the six Mn-O bonds of $MnO_6$ octahedra should also be equal at local level. Hence we tried *6 model* consisting of only one shell with six equal Mn-O bond lengths. Many workers [16,17,20] have reported the distorted $MnO_6$ octahedra at local level, which is similar to the distortion of $MnO_6$ octahedra in $LaMnO_3$. Structure of $LaMnO_3$ is orthorhombic with four Mn-O bonds in the basal plane and two in the apical plane. While in the apical plane the Mn-O bond length is 2.17 Å, in the basal plane these are 1.91 Å and 1.97 Å [26]. The long and short bond lengths differ by 0.26 Å. However, the difference between the two short Mn-O bond lengths is just 0.06 Å. Moreover, it is expected that the distortion of $MnO_6$ octahedra would reduce with the doping of Sr. Therefore, we also modeled local structure using *4+2 model* consisting of two-shells with four equal short bond lengths and two equal long bond lengths. This allows us to remain within the limits of our experimental EXAFS resolution, $\Delta R = \pi/2\ (k_{max} - k_{min})$, for the k-range used for fitting. According to our XRD results for compositions $0.5 \leq x \leq 0.8$ the structure is orthorhombic. In these compounds the bond lengths of two Mn-O bonds in the basal plane are almost same as that of the two Mn-O bonds in the apical plane but the bond lengths of the rest two bonds in the basal plane differ from them [23]. Hence in addition we used *2+4 model*, which consists of two-shell fitting with two equal short bond lengths and four equal long bond lengths. Similarly, based on the XRD results we used *3+3 model* consisting of two-shell fitting with three equal short bond lengths and three equal long bond lengths for x = 0.9 composition.

3. Results

In figure 1, we show the variation of absorption μt for different compositions in the pre-edge, edge and XANES regions from about 25 eV below the edge to 50 eV above it. Here the edge position is defined as usual as the inflection point on the main absorption edge. It is seen to vary systematically with composition as shown in the inset to this figure. There is roughly a linear increase in the edge position as the percentage of divalent Sr ions replacing the trivalent La ions increases. Similar variation in the edge position was observed in other manganite systems [20,27]. The pre-edge structure shows two peaks marked by 1 and 2 in figure 1 and the first peak on the high energy side of the absorption edge is marked as 3. These features are similar for all the samples. Simulation studies have shown that these features are sensitive to the lattice distortion [28]. If the lattice distortion decreases, intensity of these features increase. We observed the increased intensity of peaks 1, 2 and 3 with increase in x for compounds up to x = 0.8 compositions indicating that lattice distortion decrease with x. For the end compound it decreases indicating that layered type structure is more distorted than the perovskite structure. Details of pre-edge and XANES structure will be discussed elsewhere.

Figure 2 shows the $k^2$ weighted EXAFS spectra $\chi(k)$ for all the compounds. The spectra are highly structured as expected from the powder crystal nature of the samples. These spectra reveal systematic variations in peak shapes and intensities with x, indicating changes in the local structure around Mn sites. These changes are even more prominent when the structure changes from perovskite to layered type at x = 0.9, thus hinting at different local environment for Mn sites in x = 0.9 composition. The FT of these $k^2\chi(k)$ spectra taken between 3 and 13 Å$^{-1}$ is shown in figure 3. These spectra are uncorrected for the central and back-scattered phase shifts. We observe a main peak, marked 1 in the figure, which is ascribed to the first coordination shell of the central Mn atom comprising of the O-atoms of the MnO$_6$ octahedron. Further peaks corresponding to successive coordination shells are also seen. Peak 1′ close to the peak 1 for x = 0.9 composition can be identified with layered type structure. Our XRD work [23] reveals that this compound is layered type with hexagonal phase possessing 6 layered with stacking sequence of ABCACB type. In this compound Mn is surrounded by six oxygen atoms at average distance of 1.9 Å and also by one Mn atom at a distance of about 2.5 Å. Hence, peak 1′ arises from this neighbouring Mn atom. The intensity of the first peak for x = 0.1 composition is lowest, indicating that its MnO$_6$ octahedron is most distorted. With increase in the value of x the first peak becomes sharper indicating reduced distortion of the octahedra with doping.

In figures 4, 5 and 6 representative fits for x = 0.2, 0.4 and 0.6 compositions are shown. It is evident from these figures that *6 model* does not describe the experimental data at higher k values for any of the compositions. This would imply that all the Mn-O bonds are not equal and MnO$_6$ octahedron is distorted in these compounds. Between different two shell-fits, *4+2* model gives the best fit for x = 0.1, 0.2, 0.3 compositions,

*3+3 model* for x = 0.9 composition and *2+4 model* for rest of the compounds. Figure 5 shows the fits obtained using different models for x = 0.4 compound. It is clearly seen from this figure that *2+4 model* describes the experimental spectrum better in the whole k-range used for the fitting in comparison to *4+2 model*. One shell-fit gives undistorted $MnO_6$ octahedron with only one Mn-O bond distance with large Debye-Waller factors whereas two shell-fit gives distorted octahedron with two Mn-O bond distances, with one small and one large Debye-Waller factors for $0.1 \leq x \leq 0.3$ compounds and with two small Debye-Waller factors for rest of the compounds, Table 1. Shibata *et al.* [20] and Subias *et al.* [29] who carried out EXAFS studies on $La_{1-x}Sr_xMnO_3$ ($0 \leq x \leq 0.475$) and $La_{1-x}Ca_xMnO_3$ systems, respectively also observed similar behavior.

Figure 7 shows the Debye-Waller factors obtained from two shell-fits for all the compounds. In this figure $\sigma_1^2$ and $\sigma_2^2$ are Debye-Waller factors corresponding to short and long bonds, respectively. For $0.1 \leq x \leq 0.3$ compositions $\sigma_1^2$ are small and $\sigma_2^2$ are large, whereas for rest of the compounds opposite behavior is observed. While for $x \geq 0.4$ no appreciable change is observed in the values of $\sigma_1^2$ and $\sigma_2^2$, for $x \leq 0.3$ the $\sigma_1^2$ decreases and $\sigma_2^2$ increases with x. At x = 0.3, the value of $\sigma_2^2$ is 0.0194 $Å^2$. Such a large value of Debye-Waller factor indicates that there may be a large distribution in long bond lengths in the compound x = 0.3. This may result from the possibility that the number of nearest neighbours (NN) with short bond lengths is smaller than 4 and that with the long bond lengths is greater than 2 as a consequence of a combination of *4+2* and *2+4 model*. We checked this possibility explicitly by varying the number of NN with short and long bond lengths while keeping the total number of NN fixed at six. We find that the fit improves considerably when the number of short and long bonds is 2.8 and 3.2, respectively with corresponding Debye-Waller factors as 0.0098 $Å^2$ and 0.0012 $Å^2$. The values of bond lengths for the short and long bonds are found to be 1.907 Å and 1.946 Å. In the light of this result, we also checked if the quality of fit improves for x = 0.2 and 0.4 compounds if the number of NN with short or long bond lengths are varied keeping the total number of NN fixed at six. We find that the best fits are obtained only for structures of $MnO_6$ octahedron with 4 short, 2 long and 2 short, 4 long bond lengths for x = 0.2 and 0.4 compounds, respectively. These results indicate that for some composition between x = 0.2 and 0.3, a deviation from the *4+2 model* starts and completely changes to *2+4 model* for 0.3 < x < 0.4.  This would make the transition from *4+2 model* (with 4 short and 2 long bonds) to *2+4 model* (with 2 short and 4 long bonds) a smooth one

Mn-O bond lengths obtained from EXAFS fitting using different models are summarized in figure 8.  As mentioned above up to x = 0.3 compositions *4+2 model* gives the best fit. In this model we have taken four short Mn-O bond distances in basal plane and rest two long Mn-O bond distances in apical plane of the $MnO_6$ octahedron. With increase in x, the bond length in the apical plane decreases and that in the basal plane increases. For 0.4 ≤ x ≤ 0.8 compounds *2+4 model* gives the best fit. In this model two short Mn-O bond distances are in basal plane and out of four long Mn-O bond distances, two are in the basal plane and the other two are in the apical plane. For these compounds both types of bond distances decrease with increase in x and show almost linear behaviour. The two lines connecting the points of small and large bond distances are almost parallel. In x = 0.9 composition *3+3 model* gives the best fit to the data. In this model we have considered three short Mn-O bond distances in the basal plane and out of three long Mn-O bond distances, one is in basal plane and remaining two are in apical plane. However, the short and long Mn-O bond lengths differ by ~ 0.03 Å and thus the single shell model should also be adequate. The single shell model gives a fit with R-factor of 0.008, which is somewhat larger than the R-factor of 0.002 for 3+3 model. This difference in R-factor may be due to use of larger number of parameters used in fitting for the 3+3 model, as they give extra freedom to adjust the parameters to get best fit. In the 3+3 model we have two distances and two Debye-Waller factors as opposed to only one distance and one Debye-Waller factor for single shell model. It may be seen that the Debye-Waller factor and bond length for the single shell model are $4.65 \times 10^{-3}$ Å$^2$ and 1.905 Å respectively, which are in between the two Debye-Waller factors and two bond distances for the 3+3 model, Table 1. Thus the 3+3 model seems a better fit only from the R-factor point of view.

## 4. Discussion

The X-ray diffraction studies have clearly shown that our compounds $La_{1-x}Sr_xMnO_{3+\delta}$ are rhombohedral ($0.1 \leq x \leq 0.4$), orthorhombic ($0.5 \leq x \leq 0.8$), and layered type (x = 0.9) in structure [23]. EXAFS studies reveal that all the compounds show distortion in the $MnO_6$ octahedron. On general grounds if one has double exchange behaviour in these compounds such distortion should not appear in the rhombohedral structures, although it may exist in the tetragonal and orthorhombic structures. Such distortion in all structures including rhombohedral may be understood in terms of two processes: (i) size effect, including the presence of oxygen non-stoichiometry and (ii) electronic effect due to changed valency of the dopant ion. A combination of these can be interpreted in terms of electron-phonon coupling and electron polarization causing in the limit electron self trapping and polaronic effects [11,30].

The Mn-O bond distances are varying in three different fashions over the entire composition. In the first region ($0.1 \leq x \leq 0.3$) the *4+2 model* fits better, in the second region (($0.4 \leq x \leq 0.8$) *2+4 model* and finally in the third region (x = 0.9) *3+3 model*. We now discuss these three regions.

**Region-1:** compounds up to x = 0.3 composition lie in this region. Here the Mn-O bond distances in the apical plane decrease and those in the basal plane increase with composition, figure 8. The difference in these two bond lengths for x = 0.1 composition is largest, 0.19 Å and that for x = 0.3 composition is smallest, 0.08 Å, Table 1. The observed XRD structure of the compounds up to x = 0.4 is rhombohedral having all six Mn-O bond distances in $MnO_6$ octahedra equal [23]. The distribution in the Mn-O bond lengths thus indicates that the $MnO_6$ octahedra are distorted in these compounds. Presence of such a distortion at local level may be attributed to the formation of polaron in consonance with earlier work on manganites [16,17,20,31]. Louca *et al.* [16] have attributed this kind of distortion to the formation of Jahn-Teller (JT) polaron. The theoretical studies reveal existence of distortion with volume of $MnO_6$ octahedra remaining the same to be consistent with JT polaron formation. Louca *et al.* have modeled their experimental neutron diffraction data in such a way that the volume of the $MnO_6$ octahedron remains same but the number of short and long bonds vary as x changes. In our case we have fixed the number of short bonds to 4 and that of long bonds to 2 and varied the both type of bond lengths. We did not get the same volume for all the three compounds. This deviation from volume preserving behavior may be due to the contribution of ionic size effect in creating distortion in the octahedra.

**Region-2:** The compounds $0.4 \leq x \leq 0.8$ lie in this region. Here the variation in the bond distances shows different behaviour to that in the region-1. Both short and long bond distances decrease with increase in x, indicating that the volume of the $MnO_6$ octahedron decreases with x. Such a behavior cannot be identified with the formation of Jahn-Teller polaron. This distortion at local level thus may be thought of as breathing type. Since the structure for $x = 0.4$ is rhombohedral, the presence of long and short bond lengths for this composition gives indication of polaron formation at local level. The structure of the remaining compounds is orthorhombic, and it is expected to show distribution in Mn-O bond distances. There is a clear difference in the average bond length estimated from XRD [23] and the EXAFS results for the bond lengths, Table 1. The observed decrease of bond length with increase in x may occur due to ionic size effect since replacing $La^{3+}$ by $Sr^{2+}$ ion makes tolerance factor closer to unity.

From figure 8 it is seen that there is a qualitative change in the behaviour of Mn-O bond length as x increases beyond 0.3. The XRD studies reveal a structural transformation from rhombohedral to orthorhombic in the composition range $0.4 \leq x \leq 0.5$. Further, as mentioned in the results section, the transition from the 4+2 model starts for some composition between $0.2 < x < 0.3$ and completes for x between 0.3 and 0.4. One may thus conclude that local changes in $MnO_6$ octahedra as a function of x start well before global structure changes and this transformation is not an abrupt one but occurs smoothly.

**(3) Region-3:** The last compound (i.e. $x = 0.9$) lies in this region. The average structure of this compound is of layered type with hexagonal phase possessing 6 layers with stacking sequence of ABCACB type [23]. The EXAFS analysis for this composition revealed that $MnO_6$ octahedra in this compound have three short and three long Mn-O bonds in agreement with the XRD results.

## 5. Conclusions

The room temperature EXAFS studies were carried out at Mn K-edge for the complete series of $La_{1-x}Sr_xMnO_{3+\delta}$, x = 0.1 to 0.9. Detailed analysis of EXAFS spectra revealed that local structure of $MnO_6$ octahedron can be described by two long Mn-O bonds in the apical plane and four short bonds in the basal plane for compounds with x up to 0.3. However, for $0.4 \leq x \leq 0.8$ compositions it was found that there were two short Mn-O bonds in basal plane and out of remaining four long bonds; two were in the basal plane and other two in the apical plane. The observed distribution in the Mn-O bond lengths shows that $MnO_6$ octahedra are distorted for all the compounds and the local structure is different from the average one, especially for $x \leq 0.4$. The Debye Waller factors also showed cross over behavior at x = 0.4. The change in the behaviour of long and short Mn-O bond lengths at x = 0.4 is interpreted as the change in nature of distortions from Jahn-Teller type to breathing type at x = 0.4.

**Acknowledgements**
RB and SKP thank UGC-DAE CSR for financial support. AK thanks CSIR, Government of India for senior research associate position (pool scheme).

**Figure Captions**

**Figure 1**  Normalized pre-edge and XANES spectra at the Mn K-edge of $La_{1-x}Sr_xMnO_{3+\delta}$ ($0.1 \leq x \leq 0.9$). Inset shows the chemical shift of these compounds with doping with respect to x = 0.1 composition.

**Figure 2**  $k^2$- weighted EXAFS spectra $\chi(k)$ of $La_{1-x}Sr_xMnO_{3+\delta}$ ($0.1 \leq x \leq 0.9$).

**Figure 3**  Fourier Transform of $k^2\chi(k)$ for $La_{1-x}Sr_xMnO_{3+\delta}$ ($0.1 \leq x \leq 0.9$).

**Figure 4**  Fitted patterns of $La_{0.8}Sr_{0.2}MnO_{3+\delta}$ compound using *6 model* having six equal Mn-O bonds and *4+2 model* having four short and two long Mn-O bonds.

**Figure 5**  Fitted patterns of $La_{0.6}Sr_{0.4}MnO_{3+\delta}$ compound using *6 model* having six equal Mn-O bonds, *4+2 model* having four short and two long Mn-O bonds and *2+4 model* having two short and four long Mn-O bonds.

**Figure 6**  Fitted patterns of $La_{0.4}Sr_{0.6}MnO_{3+\delta}$ compound using *6 model* having six equal Mn-O bonds and *2+4 model* having two short and four long Mn-O bonds.

**Figure 7**  Debye-Waller factors $\sigma_1^2$ and $\sigma_2^2$ corresponding to short and long bonds, respectively obtained from two-shell fitting for $La_{1-x}Sr_xMnO_{3+\delta}$ ($0.1 \leq x \leq 0.9$). The values of Debye-Waller factors for $0.1 \leq x \leq 0.3$, $0.4 \leq x \leq 0.8$ and x = 0.9 correspond to *4+2 model*, *2+4 model* and *3+3 model*, respectively. For details please see the text.

**Figure 8**  Short bond length ($R_1$) and long bond length ($R_2$) obtained from two-shell fitting for $La_{1-x}Sr_xMnO_{3+\delta}$ ($0.1 \leq x \leq 0.9$).

**Table Captions**

**Table 1**  Oxygen non-stoichiometry δ obtained from iodometric redox titration and parameters obtained from EXAFS fitting of $La_{1-x}Sr_xMnO_{3+\delta}$ ($0.1 \leq x \leq 0.9$). $R_1$ and $R_2$ are short and long Mn-O bond lengths, respectively. $\sigma_1^2$ and $\sigma_2^2$ are Debye-Waller factors corresponding to short and long bonds, respectively. R-factor is the goodness of fit parameter. The values of $R_1$, $R_2$, $\sigma_1^2$ and $\sigma_2^2$ for $0.1 \leq x \leq 0.3$, $0.4 \leq x \leq 0.8$ and $x = 0.9$ correspond to *4+2 model*, *2+4 model* and *3+3 model*, respectively. For details please see the text.

**Table 1**

| $La_{1-x}Sr_xMnO_{3+\delta}$ | $\delta$ | $R_1$ (Å) | $R_2$ (Å) | $\sigma_1^2$ ($\times 10^{-3}$ Å$^2$) | $\sigma_2^2$ ($\times 10^{-3}$ Å$^2$) | R-factor |
|---|---|---|---|---|---|---|
| x = 0.1 | 0.041(1) | 1.935(3) | 2.133(7) | 3.16(28) | 6.40(71) | 0.0020 |
| x = 0.2 | 0.036(3) | 1.939(8) | 2.066(9) | 2.81(54) | 11.1(17) | 0.0026 |
| x = 0.3 | 0.0231(1) | 1.945(7) | 2.022(12) | 2.33(40) | 19.4(22) | 0.0076 |
| x = 0.4 | 0.0162(5) | 1.872(7) | 1.956(7) | 4.17(1.45) | 2.56(58) | 0.0019 |
| x = 0.5 | 0.012(1) | 1.866(5) | 1.945(5) | 4.2(1.0) | 2.56(40) | 0.0013 |
| x = 0.6 | 0.011(1) | 1.856(7) | 1.938(7) | 4.2(1.4) | 2.56(56) | 0.0023 |
| x = 0.7 | 0.013(1) | 1.852(5) | 1.933(5) | 4.01(55) | 2.62(22) | 0.0023 |
| x = 0.8 | 0.017(6) | 1.851(12) | 1.925(8) | 4.2(1.7) | 2.83(44) | 0.0034 |
| x = 0.9 | -0.005(6) | 1.89(4) | 1.917(2) | 6.73(1.65) | 2.15(50) | 0.0019 |

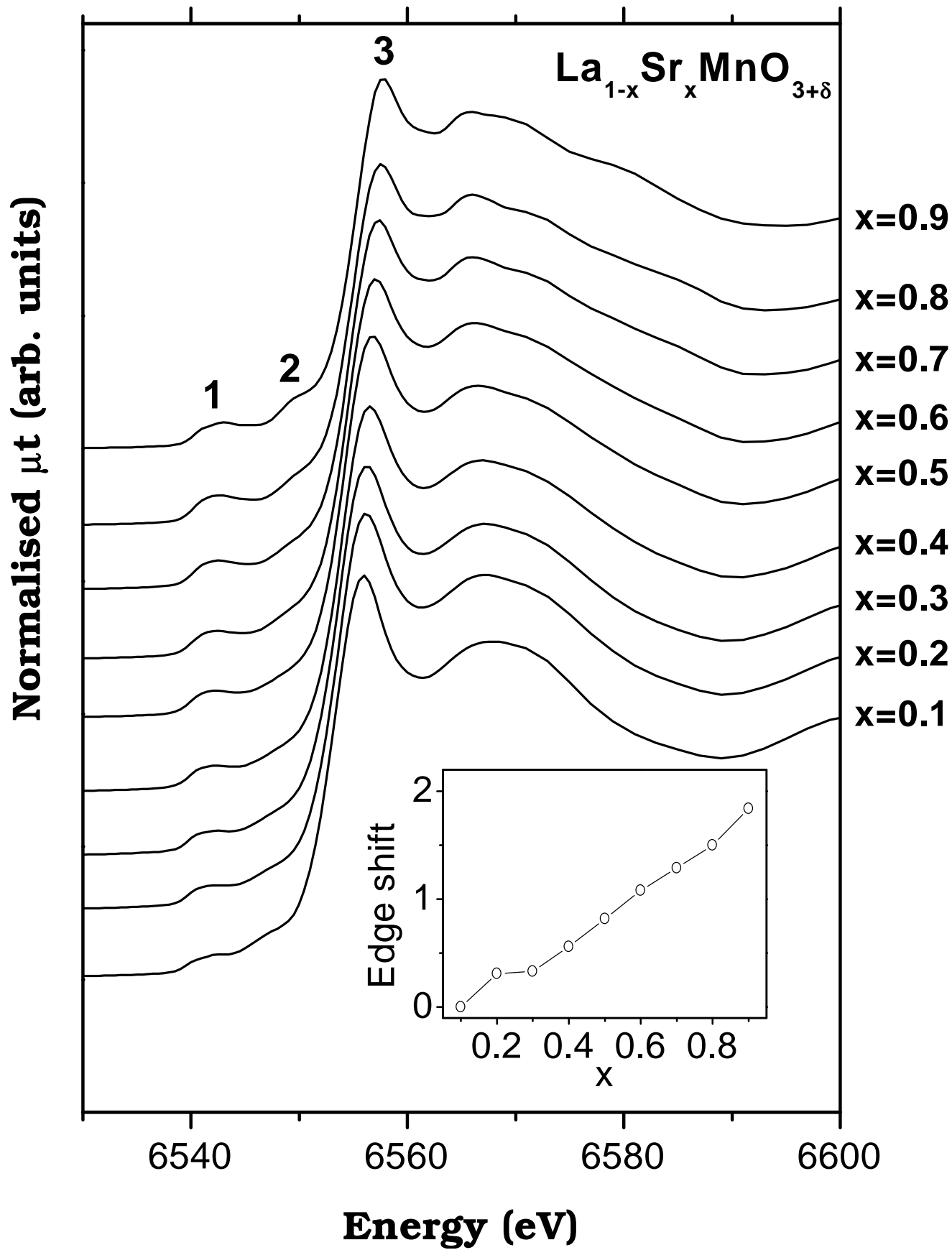

Figure 1
Bindu et al.

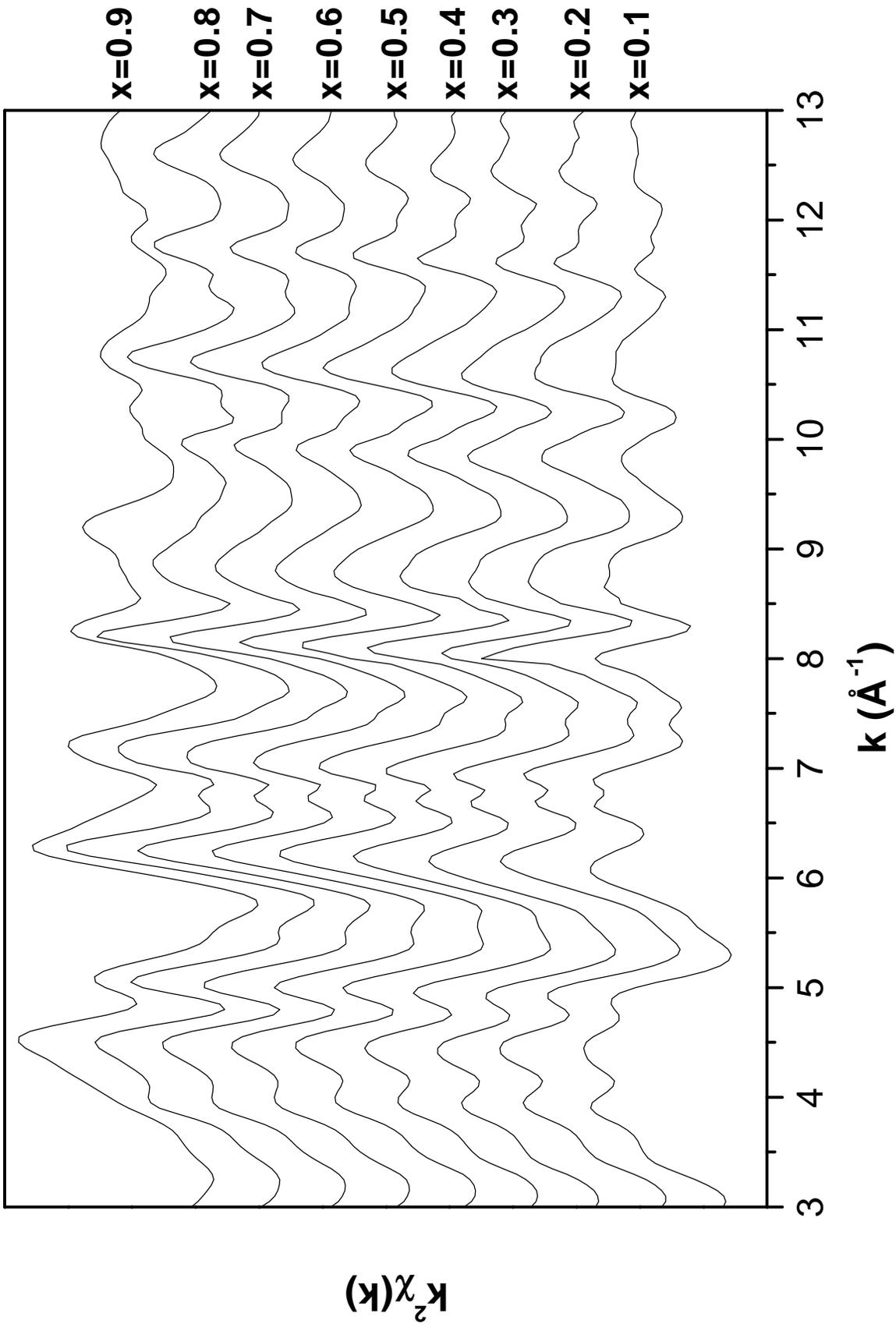

Figure 2
Bindu et al.

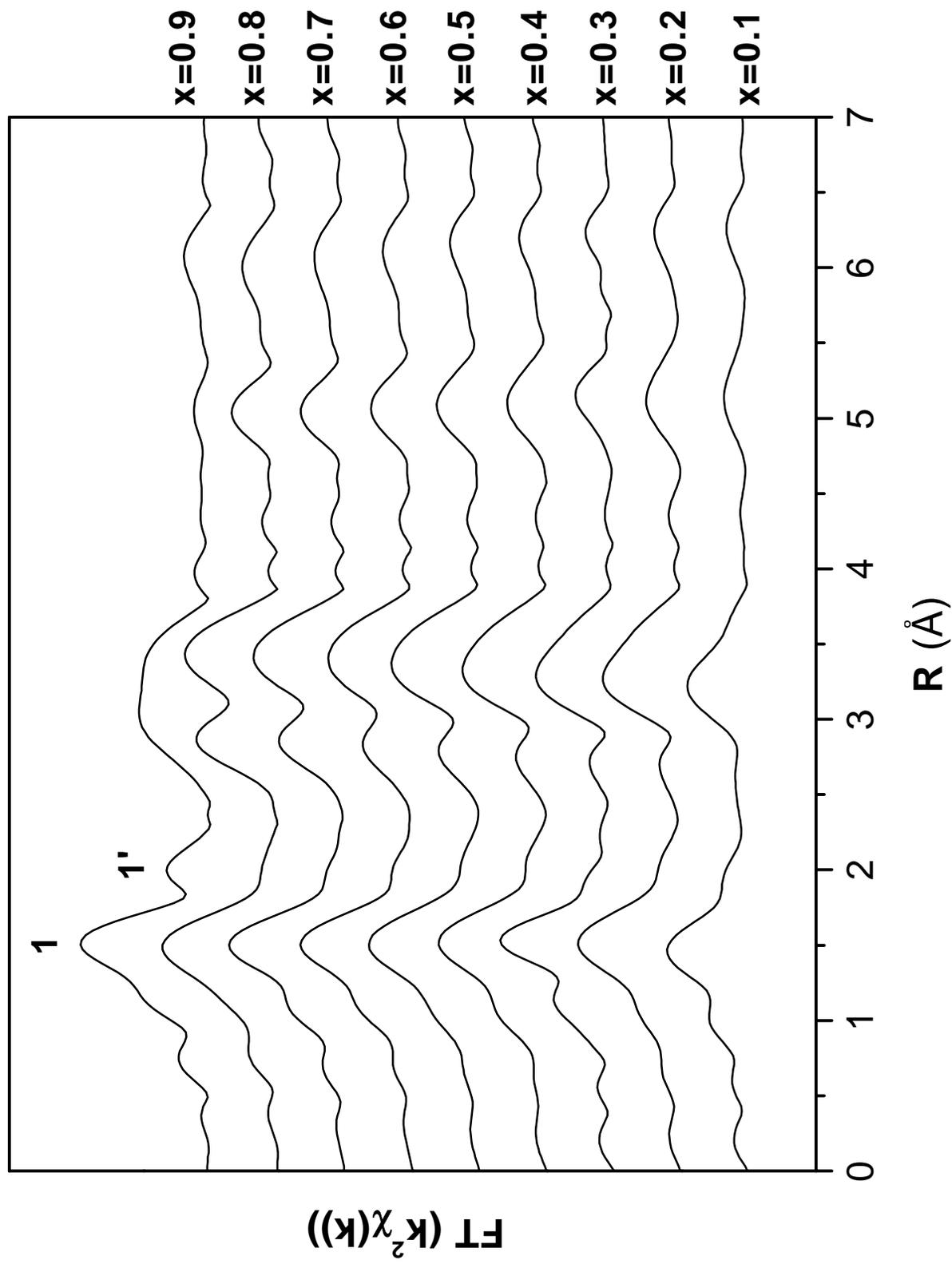

Figure 3
Bindu et al.

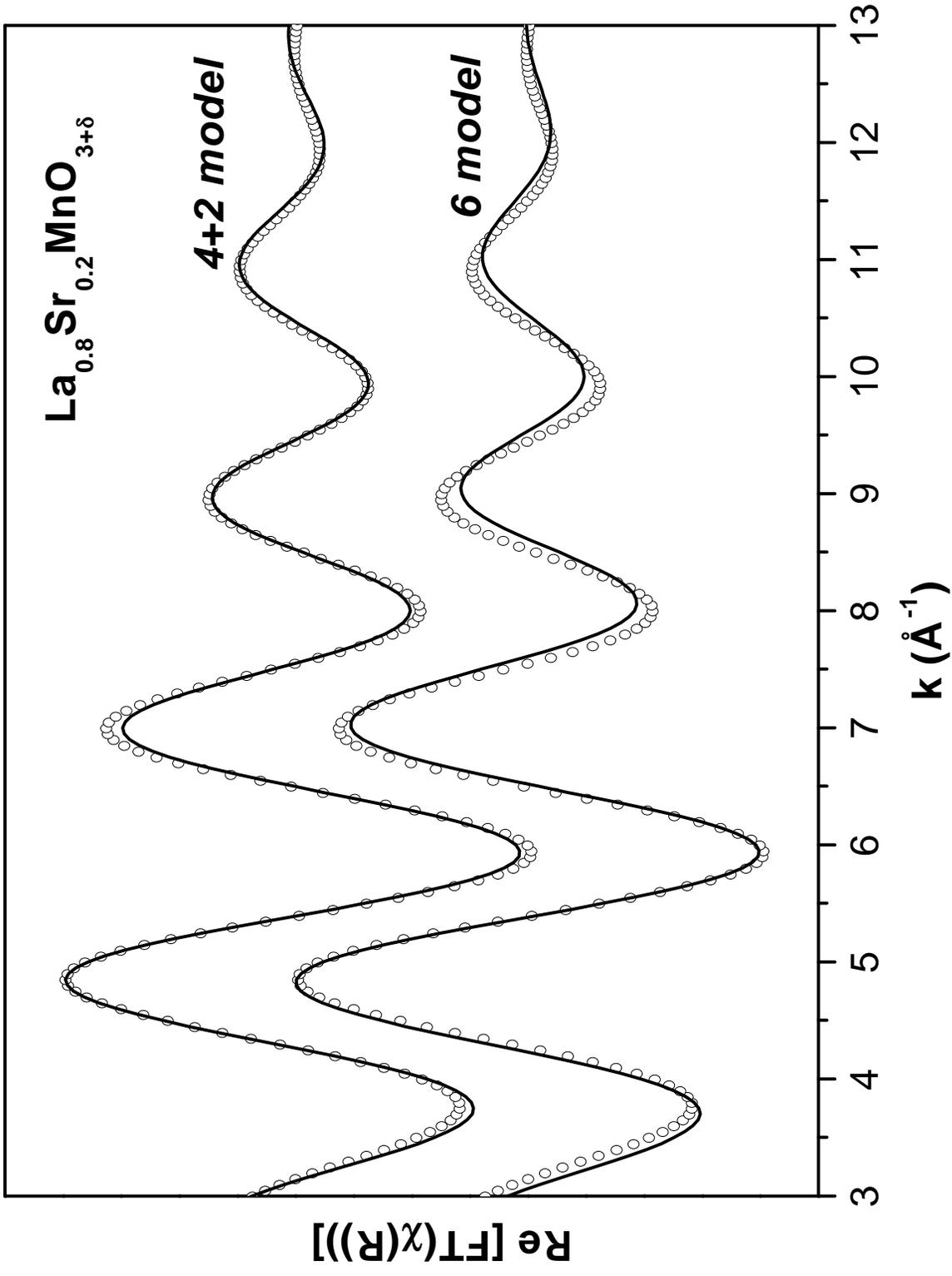

Figure 4
Bindu et al.

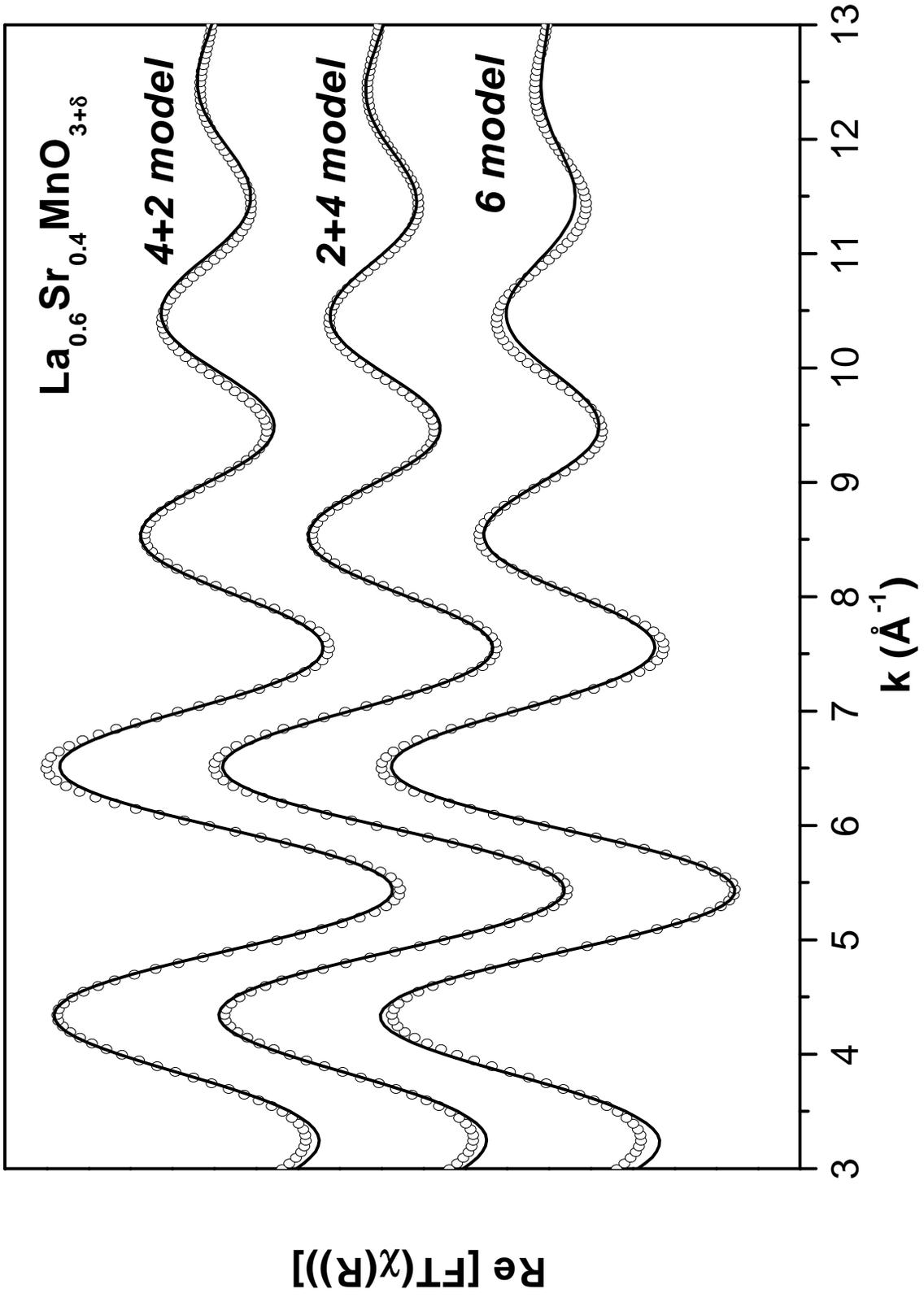

Figure 5
Bindu et al.

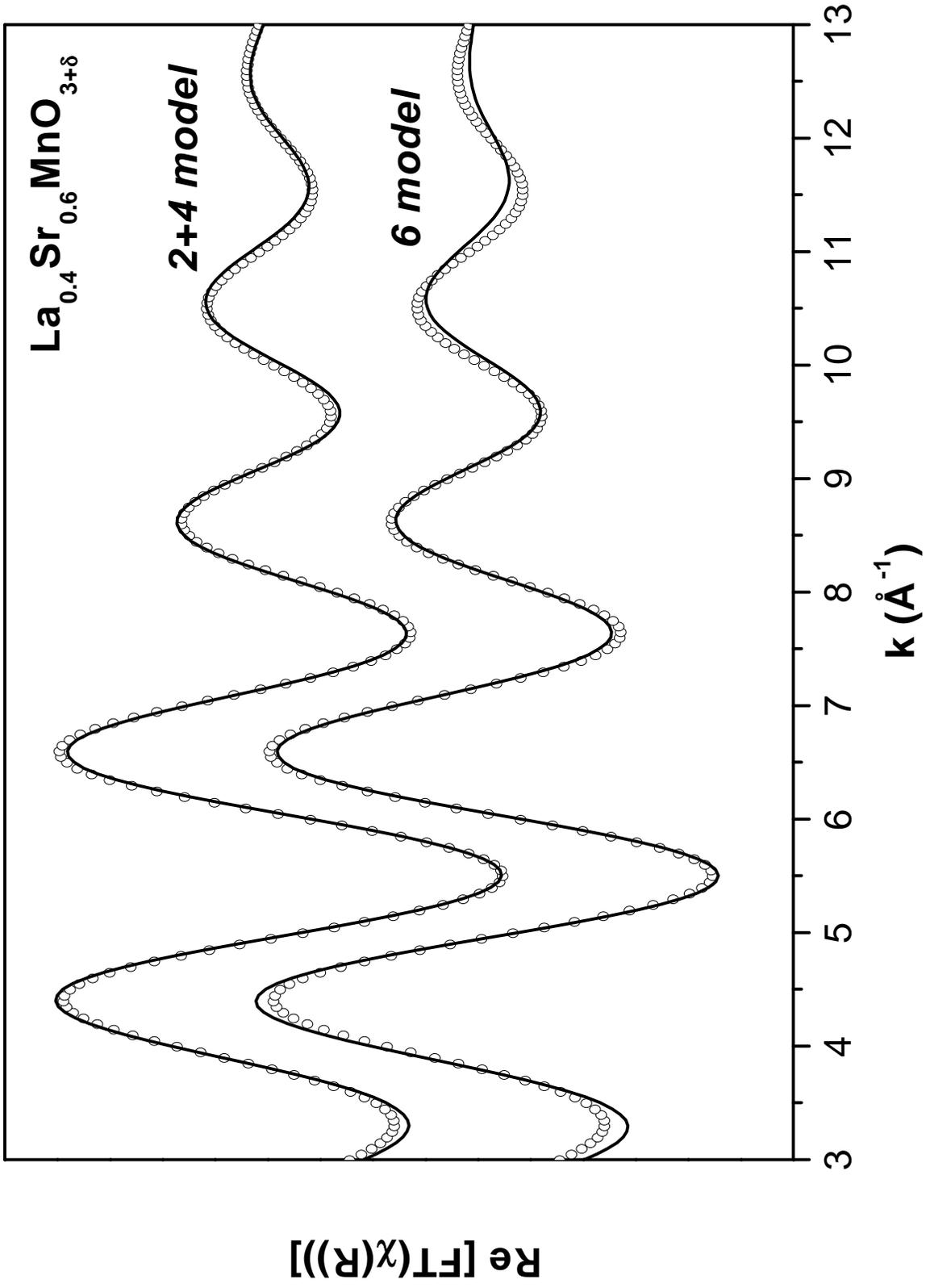

Figure 6
Bindu et al.

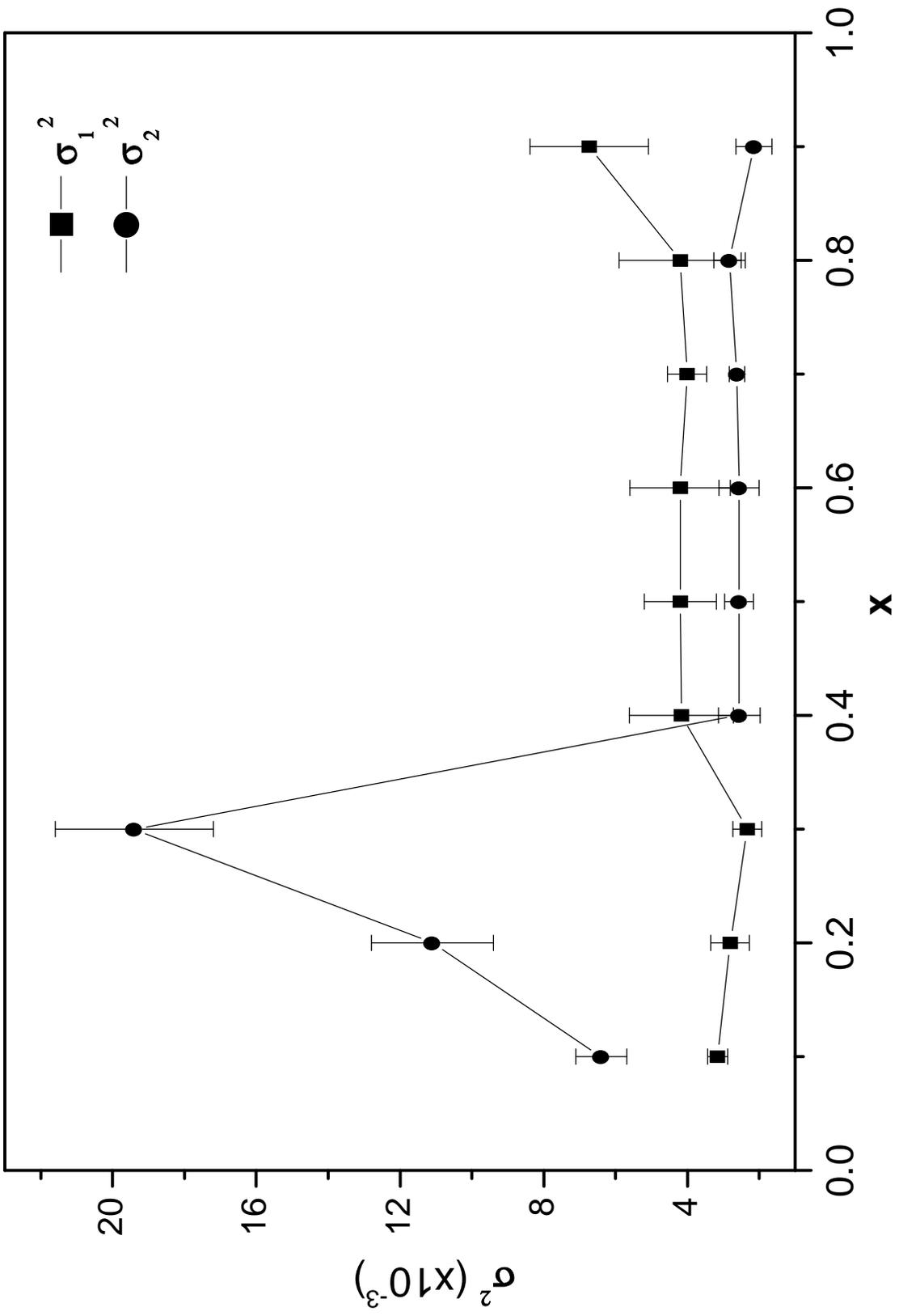

Figure 7
Bindu et al.

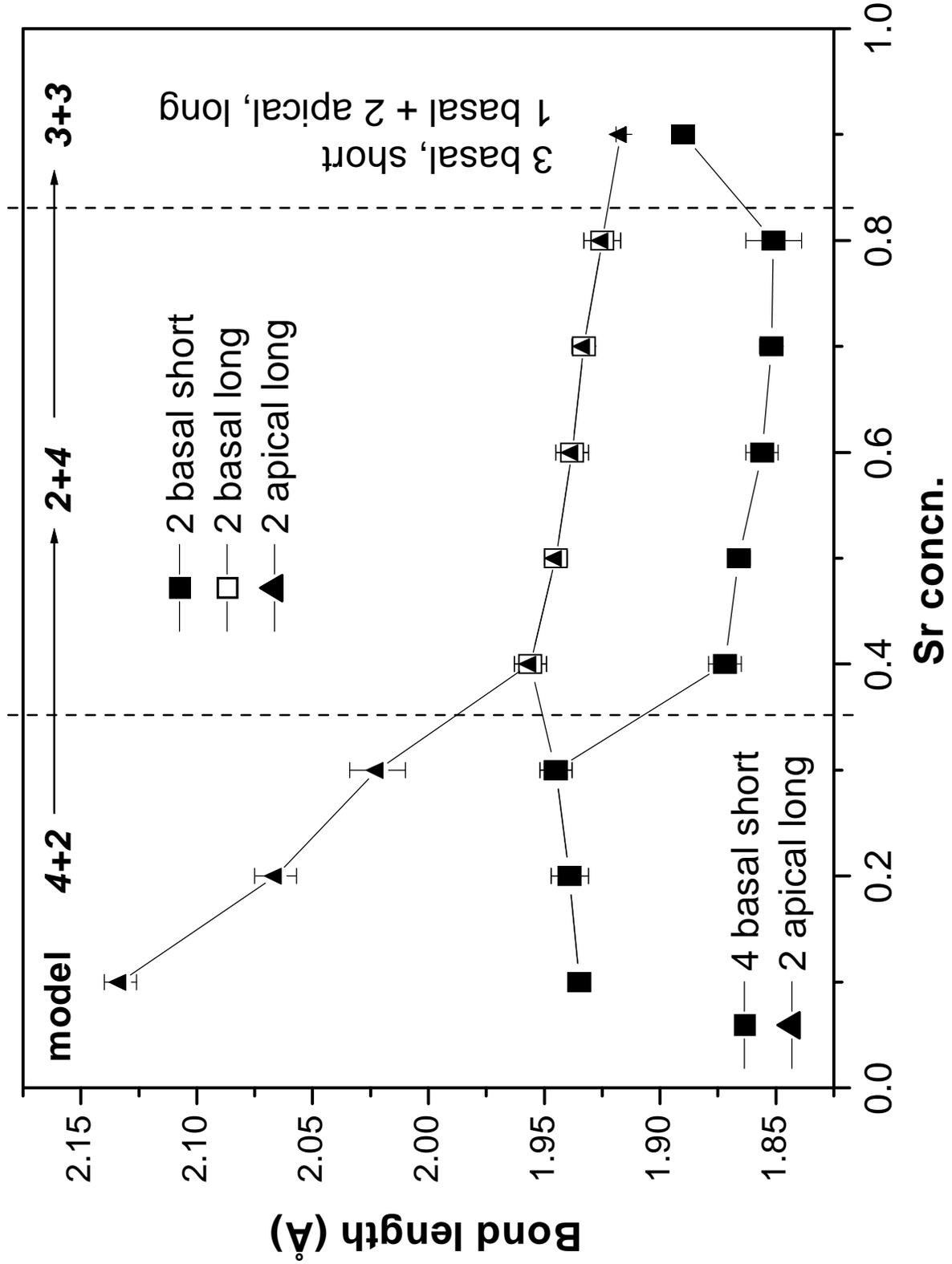

Figure 8
Bindu et al.